\documentstyle[preprint,aps,epsf,graphicx]{revtex}

\begin{document}

\tightenlines

\title{ $K^-$/$K^+$ ratio in heavy-ion collisions at GSI with an
  antikaon self-energy in hot and dense matter } \author{Laura
  Tol\'os, Artur Polls, Angels Ramos} \address{Departament
  d'Estructura i Constituents de la Mat\`eria,
  Universitat de Barcelona, \\
  Diagonal 647, 08028 Barcelona, Spain }

\author{J\"urgen Schaffner-Bielich} 

\address{Institut f\"ur Theoretische Physik,
J. W. Goethe-Universit\"at\\
D-60054 Frankfurt am Main, Germany}

\date{\today}

\maketitle

\begin{abstract}
  The $K^-$/$K^+$ ratio produced in heavy-ion collisions at GSI
  energies is studied. The in-medium properties at finite temperature
  of the hadrons involved are included, paying a special attention to
  the in-medium properties of antikaons.
  Using a statistical approach, it is found that the determination of
  the temperature and chemical potential at freeze-out conditions
  compatible with the ratio $K^-$/$K^+$ is very delicate,
  and depends very strongly on the approximation adopted for the
  antikaon self-energy. The use of an energy dependent $\bar{K}$
  spectral density, including both s and p-wave components of the
  $\bar{K}N$ interaction, lowers substantially the freeze-out
  temperature compared to the standard simplified mean-field treatment
  and gives rise to an overabundance of $K^-$ production in the dense
  and hot medium. Even a moderately attractive antikaon-nucleus
  potential obtained from our self-consistent many-body calculation
  does reproduce the ``broad-band equilibration'' advocated by Brown,
  Rho and Song due to the additional strength of the spectral function
  of the $K^-$ at low energies.

  
  \vspace{0.5cm}



\end{abstract}


\section{Introduction}
\label{sec:intro}

The study of the properties of hadrons in hot and dense matter is
receiving a lot of attention in recent years to understand fundamental
aspects of the strong interaction, such as the partial restoration of
chiral symmetry \cite{brown-rho02,wambach,mosel}, as well as a variety
of astrophysical phenomena, such as the dynamical evolution of
supernovas and the composition of neutron stars \cite{Heiselberg97}.

A special effort has been invested to understand the properties of
antikaons in the medium, especially after the speculation of the
possible existence of an antikaon condensed phase was put forward in
\cite{KN86} which would soften the equation-of-state producing, among
other phenomena, lower neutron star masses \cite{Brown94,Li97a,kaon}.

While it is well established that the antikaons should feel an
attractive interaction when they are embedded in a nuclear
environment, the size of this attraction has been the subject of
intense debate. Theoretical models including medium effects on the
antikaon-nucleon scattering amplitude, the behaviour of which is
governed by the isospin zero $\Lambda(1405)$ resonance, naturally
explain the evolution from repulsion in free-space to attraction in
the nuclear medium \cite{koch,weise}. However, the presence of the
resonance makes the size of the antikaon-nucleus potential be very
sensitive to the many-body treatment of the medium effects.  The
phenomenological attempts of extracting information about the
antikaon-nucleus potential from kaonic-atom data favoured very
strongly attractive well depths \cite{FGB94}, but recent
self-consistent in-medium calculations
\cite{Lutz98,Oset00,Schaffner,Laura01} based on chiral Lagrangian's
\cite{weise,oset98} or meson-exchange potentials \cite{holinde90}
predict a moderately attractive kaon-nucleus interaction. In fact,
recent analysis of kaonic atoms \cite{satoru00,baca,cieply01} are able
to find a reasonable reproduction of the data with relative shallow
antikaon-nucleus potentials, therefore indicating that kaonic atom
data cannot really pin down the strength of the antikaon-nucleus
potential at nuclear matter saturation density.

On the other hand, heavy-ion collisions at energies around $1-2$ AGeV
offer the possibility of studying experimentally the properties of a
dense and hot nuclear system \cite{Oeschler02,Senger01,Sturm02}. In
particular, a considerable amount of information about strange
particles like antikaons is available. Since antikaons are produced at
finite density and finite momentum, the chiral models have recently
incorporated the higher partial waves of the antikaon-nucleon
scattering amplitude both in free space\cite{caro,Lutz021,Jido} and in
the nuclear medium \cite{Lutz022,Kolomeitsev02}. The complete scenario
taking into account finite density, finite momentum and finite
temperature has recently been addressed in Refs.\ 
\cite{Schaffner,Laura02}.

Event generators trying to analyse heavy-ion collision data
\cite{ko87,teis97,effenberg99,fuchs01,hartnack98,cassing97,sibirt98}
need to implement the modified properties of the hadrons in the medium
where they are produced. Transport models have shown, for instance,
that the multiplicity distributions of kaons and antikaons are much
better reproduced if in-medium masses rather than bare masses are used
\cite{Li98,cas-elena}. Production and propagation of kaons and
antikaons have been investigated with the Kaon Spectrometer (KaoS) of
the SIS heavy-ion synchrotron at GSI (Darmstadt). The experiments have
been performed with Au$+$Au, Ni$+$Ni, C$+$C at energies between $0.6$
and $2.0$ AGeV
\cite{miskowiec,ahner,shin,kaos97,kaos99,menzel,sturm,ritman,best,crochet}.
One surprising observation in C$+$C and Ni$+$Ni collisions
\cite{kaos97,kaos99,menzel} is that, as a function of the energy
difference $\sqrt s-\sqrt s_{th}$, where $\sqrt s_{th}$ is the needed
energy to produce the particle ($2.5$ GeV for $K^+$ via $pp
\rightarrow \Lambda K^+ p$ and $2.9$ GeV for $K^-$ via $pp \rightarrow
pp K^- K^+ $), the number of $K^-$ balanced the number of $K^+$ in
spite of the fact that in $pp$ collisions the production
cross-sections close to threshold are 2-3 orders of magnitude
different. This has been interpreted to be a manifestation of the
enhancement of the $K^+$ mass and the reduction of the $K^-$ one in
the nuclear medium, which in turn influence the corresponding
production thresholds \cite{Li97a,cassing97,sibirt98,Li98,cas-elena},
although a complementary explanation in terms of in-medium enhanced
$\pi\Sigma \to K^- p$ production has also been given \cite{Schaffner}.
Another interesting observation is that at incident energies of $1.8$
and $1.93$ AGeV, the $K^-$ and $K^+$ multiplicities have the same
impact parameter dependence \cite{kaos97,kaos99,menzel}. Equal
centrality dependence for $K^+$ and $K^-$ and, hence, independence of
centrality for the $K^-/K^+$ ratio has also been observed in Au$+$Au
and Pb$+$Pb reactions between $1.5$ AGeV and RHIC energies
($\sqrt{s}=200$ AGeV) \cite{menzel,forster,ahle,harris,dunlop}. This
independence of centrality is astonishing, since at low energies one
expects that as centrality increases --- and with it the participating
system size and the density probed --- the $K^-/K^+$ ratio should also
increase due to the increased reduction of the $K^-$ mass together
with the enhancement of the $K^+$ mass. In fact, the independence of
the $K^-/K^+$ ratio on centrality has often been advocated as
signalling the lack of in-medium effects. A recent interesting
interpretation of this phenomenon is given in Ref.~\cite{Hartnack},
where it is shown that the $K^-$ are predominantly produced via $\pi
Y$ collisions ($Y=\Lambda,\Sigma$) and, hence, the $K^-$ multiplicity
is strongly correlated with the $K^+$ one, since kaons and hyperons
are mainly produced together via the reaction $NN\to K Y N$.

Although the transport model calculations show that strangeness
equilibration requires times of the order of $40-80$ fm/c
\cite{pkoch,brato}, statistical models, which assume chemical and
thermal equilibrium and common freeze-out parameters for all
particles, are quite successful in describing particle yields
including strange particles
\cite{Cley98,Cleynucl,Cley57,Cley59,Cley60,Cley00}. The kaon and
antikaon yields in the statistical models are based on free masses and
no medium effects are needed to describe the enhanced in-medium
$K^-/K^+$ ratio or its independence with centrality.  The increased
value of the $K^-/K^+$ ratio is simply obtained by choosing a
particular set of parameters at freeze-out, the baryonic chemical
potential $\mu_B \simeq 720$ MeV and the temperature $T \simeq 70$
MeV, which also reproduce a variety of particle multiplicity ratios
\cite{Cley60,Cley00}.  On the other hand, centrality independence of
the $K^-/K^+$ ratio is automatically obtained in statistical models
within the canonical or grand-canonical schemes because the terms
depending on the system size drop out \cite{Cley00}. However, as shown
by Brown et al. in Ref.~\cite{brown01}, using the reduced in-medium
$K^-$ mass in the statistical model would force, in order to reproduce
the experimental value of the $K^-/K^+$ ratio, a larger value of the
chemical potential and hence a larger and more plausible baryonic
density for strangeness production. In addition, Brown et al.
introduce the concept of ``broad-band equilibration" according to
which the $K^-$ mesons and the hyperons are produced in an essentially
constant ratio independent of density, hence explaining also the
centrality independence of the $K^-/K^+$ ratio but including medium
effects.  In essence, the establishment of a broad-band relies on the
fact that the baryonic chemical potential $\mu_B$ increases with
density an amount which coincides roughly with the reduction of the
$K^-$ mass. However, as mentioned above, the antikaon properties are
very sensitive to the type of model for the ${\bar K}N$ interaction
used and of the in-medium effects included.  The purpose of the
present work is precisely to investigate to which extent the
``broad-band equilibration" concept holds when more sophisticated
models for the in-medium antikaon properties are used. We explore
within the context of a statistical model what is the behaviour of the
$K^-/K^+$ ratio as a function of the nuclear density when the hadrons
are dressed, paying special attention to different ways of dressing
the antikaons: either treating them as non-interacting, or dressing
them with an on-shell self-energy, or, finally, considering the
complete antikaon spectral density. We use the antikaon self-energy
which has been derived within the framework of a coupled-channel
self-consistent calculation in symmetric nuclear matter at finite
temperature \cite{Laura02}, taking as bare meson-baryon interaction
the meson exchange potential of the J\"ulich group \cite{holinde90}.
We will show that the determination of temperature and chemical
potential at freeze-out conditions compatible with the experimental
value of the $K^-$/$K^+$ ratio is very delicate, and depends very
strongly on the approximation adopted for the antikaon self-energy.

The paper is organised as follows. In Sec.~II the formalism of the
thermal model is described and the different ingredients used in the
determination of the off-shell properties of the $K^-$ are given. The
results are presented and discussed in Sec.~III. Finally, our
concluding remarks are given in Sec.~IV.

\section{Formalism}
\label{sec:formalism}

In this section we present a brief description of the thermal
models to account for strangeness production in heavy-ion
collisions. The basic hypothesis is to assume that the relative
abundance of kaons and antikaons in the final state of
relativistic nucleus-nucleus collisions is determined by imposing
thermal and chemical equilibrium
~\cite{Cley98,Cleynucl,Cley57,Cley59,Cley60,Cley00,Hagedorn}. The
fact that the number of strange particles in the final state is
small requires a strict treatment of the conservation of
strangeness and, for this quantum number, one has to work in the
canonical scheme. Other conservation laws must also be imposed,
like baryon number and electric charge conservation. Since the
number of baryons and charged particles is large, they can be
treated in the grand-canonical ensemble. In this way, the
conservation laws associated to these other quantum numbers are
satisfied on average, allowing for fluctuations around the
corresponding mean values.

To restrict the ensemble according to the exact strangeness
conservation law, as done in 
Refs.~\cite{Cley98,Cleynucl,Cley57,Cley59,Cley60,Cley00},
one has to project the grand-canonical partition function,
$\bar{Z}(T,V,\lambda_B,\lambda_S,\lambda_Q)$, onto a fixed value
of strangeness $S$,

\begin{eqnarray}
Z_{S}(T,V,\lambda_B,\lambda_Q)=\frac{1}{2\pi} \int_{0}^{2\pi} \ d\phi \ e^{-iS\phi} \
\bar{Z}(T,V,\lambda_B,\lambda_S,\lambda_Q) \ ,
\label{canon1}
\end{eqnarray}
where  $\lambda_B,\lambda_S,\lambda_Q$ are the baryon, strangeness
and charge fugacities, respectively, and where $\lambda_S$ stands
explicitly for $\lambda_S= e^{i \phi}$.

Only particles with $S=0,\pm 1$ are included in the
grand-canonical partition function  because, in the range of
energies achieved at GSI, they are produced with a higher
probability than particles with $S=\pm 2, \pm 3$. The
grand-canonical partition function is calculated assuming an
independent particle behaviour and the Boltzmann approximation for
the one-particle partition function of the different particle
species.
In principle, one deals with a dilute system, so the independent
particle model seems justified. However, medium effects on the
particle properties can be relevant. As mentioned in the
Introduction, the aim of this paper is to study how the dressing of
the hadrons present in the gas, especially the antikaons, affects
the observables such as the ratio of kaon and antikaon particle
multiplicities, in particular for the conditions of the heavy-ion
collisions at SIS/GSI energies.

Within the approximations mentioned above, the grand-canonical
partition function reads as follows,
\begin{eqnarray}
\bar{Z}(T,V,\lambda_B,\lambda_S,\lambda_Q)=
{\rm exp}(N_{S=0}+N_{S=-1}e^{-i\phi}+N_{S=1}e^{i\phi}) \ ,
\label{canon2}
\end{eqnarray}
where $N_{S=0,\pm 1}$ is the sum over  one-particle partition functions of all particles and resonances with strangeness $S=0,\pm 1$,
\begin{eqnarray}
N_{S=0,\pm 1} &=& \sum_{B_i} Z_{B_i}^1+ \sum_{M_j} Z_{M_j}^1+ \sum_{R_k} Z_{R_k}^1 ,\\
Z_{B_i}^1&=& g_{B_i} \ V \ \int \ \frac{d^3p}{(2\pi)^3} 
\ e^{\frac{-E_{B_i}}{T}} \ e^{\frac{\mu_{B_i}}{T}} \ e^{\frac{\mu_{Q(B_i)}}{T}} \ ,
\\
Z_{M_j}^1&=& g_{M_j} \ V \ \int \ \frac{d^3p}{(2\pi)^3} \ e^{\frac{-E_{M_j}}{T}}  \ e^{\frac{\mu_{Q(M_j)}}{T}} \ , \\
Z_{R_k}^1&=& g_{R_k} \ V \ \int \ \frac{d^3p}{(2\pi)^3} \int_{m-2\Gamma}^{m+2\Gamma} \ 
ds \ e^{\frac{-\sqrt{p^2+s}}{T}} \frac{1}{\pi}
\frac{m\Gamma}{(s-m^2)^2+m^2\Gamma^2}
\ \left( e^{\frac{\mu_B(R_k)}{T}} \right) \
e^{\frac{\mu_Q(R_k)}{T}}  \ .
\label{reso}
\end{eqnarray}
The expressions $Z_{B_i}^1$ and $Z_{M_j}^1$ indicate the
one-particle partition function for baryons and mesons
respectively, while $Z_{R_k}^1$ is the one-particle partition function
associated to a baryonic or mesonic resonance. In the latter case,
however, the factor $e^{\frac{\mu_B(R_k)}{T}}$ would not be
present in Eq.~(\ref{reso}). Notice that the resonance is
described by means of a Breit-Wigner parameterisation. The
quantity V is the interacting volume of the system, $g_B$, $g_M$
and $g_R$ are spin-isospin degeneracy factors and $\mu_B$ and
$\mu_Q$ are the baryonic and charge chemical potentials of the
system. For Ni+Ni system at SIS energies, $\mu_Q$ can be omitted
because it is associated to the isospin-asymmetry of the system
and, in this case, the deviation from the isospin-symmetric case
is only $4\%$ (see Ref. \cite{Cley57}). On the other hand, $\mu_B$
will be fixed to the nucleonic chemical potential, $\mu_N$,
because the abundance of nucleons is larger than the one for the other
baryons produced. The energies $E_B$, $E_M$ refer to the in-medium
single-particle energies of the hadrons present in the system at a
given temperature.

Following  Ref.~\cite{Cley57}, the canonical partition function
for total strangeness $S=0$ is

\begin{eqnarray}
Z_{S=0}(T,V,\lambda_B)=\frac{1}{2\pi} \int_{0}^{2\pi} \ d\phi  \
{\rm exp}(N_{S=0}+N_{S=-1}e^{-i\phi}+N_{S=1}e^{i\phi}) \ .
\label{canon3}
\end{eqnarray}

In this work, as well as in Ref.~\cite{Cley59}, the small and
large volume limits of the particle abundances were studied. These
limits were performed to show that the canonical treatment of
strangeness in obtaining the particle abundances gives completely
different results in comparison to the grand-canonical scheme,
demonstrating at the same time that, for the volume considered,
the canonical scheme is the appropriate one. The aim of going
through these limits again in the following is to remind the reader
that, for
the specific case of the ratio $K^-/K^+$, the result is
independent of the size of the system and is the same for both the
canonical and grand-canonical treatments.


According to statistical mechanics, to compute the number of kaons
and antikaons ~\cite{Hagedorn} one has to differentiate the
partition function with respect to the particle fugacity
\begin{eqnarray}
N_{K^-(K^+)} \equiv \left( \lambda_{K^-(K^+)}
\frac{\partial}{\partial\lambda_{K^-(K^+)}} \ln
Z_{S=0}(\lambda_{K^-(K^+)}) \right)_{\lambda_{K^-(K^+)}=1} \ .
\end{eqnarray}
Expanding $Z_{S=0}$ in the small volume limit, $N_{K^-}$ and
$N_{K^+}$ are given by
\begin{eqnarray}
N_{K^+}&=& g_{K^+}  V  \int
\frac{d^3p}{(2\pi)^3} e^{\frac{-E_{K^+}}{T}} \times \nonumber \\
&&\left\lbrace \sum_i  g_i  V  \int \frac{d^3p}{(2\pi)^3}
e^{\frac{-E_{B_i(S=-1)}+\mu_{B_i}}{T}}+
 \sum_j  g_j  V  \int \frac{d^3p}{(2\pi)^3}
 e^{\frac{-E_{M_j(S=-1)}}{T}}
+\sum_k Z_{R_{k(S=-1)}} \right\rbrace  \ ,
\nonumber \\
N_{K^-}&=& g_{K^-} V \int \frac{d^3p}{(2\pi)^3}
e^{\frac{-E_{K^-}}{T}} \times \nonumber \\
&&\left\lbrace \sum_j g_j V \int \frac{d^3p}{(2\pi)^3}
e^{\frac{-E_{M_j(S=+1)}}{T}} +\sum_k Z_{R_{k(S=+1)}}
\right\rbrace \ . \label{eq:smallvolume}
\end{eqnarray}
where antibaryons have not been considered because the ratio
$\frac{\bar{B}}{B}=e^{-\frac{2\mu_B}{T}}$ is negligibly small at
GSI/SIS colliding energies, where $B$ and $\bar{B}$ represent the
number of baryons and antibaryons, respectively. The expression
for $N_{K^+}$ ($N_{K^-}$) indicates that the number of
$K^+$($K^-$) has to be balanced with all particles and resonances
with $S=-1$($S=1$). It can be observed from
Eq.~(\ref{eq:smallvolume}) that the ratio $K^-$/$K^+ \equiv
N_{K^-}/N_{K^+}$ in the canonical ensemble does not depend on the
volume because it cancels out exactly.

At the other extreme, i.e. in the thermodynamic limit (large
volumes), since it is known that the canonical treatment is equivalent
to the grand canonical one, one can compute the ratio explicitly
from the grand canonical partition function
$\bar{Z}(T,V,\lambda_B,\lambda_S)$,

\begin{eqnarray}
\ln \bar{Z}(T,V,\lambda_B,\lambda_S)=\lambda_S
Z_{K^+}^1+\frac{1}{\lambda_S} Z_{K^-}^1+\lambda_B
\frac{1}{\lambda_S} Z_{B,S=-1}^1 +\lambda_S
Z_{M,S=+1}^1+\frac{1}{\lambda_S} Z_{M,S=-1}^1 \ ,
\end{eqnarray}
where $Z_{B,S=\pm1}^1$($Z_{M,S=\pm1}^1$) is the sum of
one-particle partition functions for baryons (mesons) with
$S=\pm1$. Then, by  imposing strangeness conservation on average,

\begin{eqnarray}
\langle S \rangle=\lambda_S \ \frac{\partial}{\partial \lambda_S} \ \ln
\bar{Z}=0,
\end{eqnarray}
one can easily obtain $\lambda_S$,
\begin{eqnarray}
\lambda_S^2=\frac{Z_{K^-}^1+\lambda_B \ Z_{B,S=-1}^1+
Z_{M,S=-1}}{Z_{K^+}^1+ Z_{M,S=+1}^1} \ .
\end{eqnarray}
Therefore, from
\begin{eqnarray}
\langle N_{K^+} \rangle= \lambda_S \ Z_{K^+}^1 \nonumber \\
\langle N_{K^-} \rangle= \frac{1}{\lambda_S} \ Z_{K^-}^1,
\end{eqnarray}
one obtains the ratio
\begin{eqnarray}
\frac{K^-}{K^+}=\frac{Z_{K^-}^1}{Z_{K^+}^1} \ \frac{Z_{K^+}^1+
Z_{M,S=+1}^1} {Z_{K^-}^1+\lambda_B \ Z_{B,S=-1}^1+ Z_{M,S=-1}} \ .
\label{eq:largevolume}
\end{eqnarray}
The condition $\langle S \rangle=0$ and
dealing with strange particles of $S=0,\pm1$ makes the ratio
independent of the volume.

Although the expression obtained is the same as that from the
canonical ensemble in the small volume limit, the proof that the
ratio $K^-/K^+$ is independent of the volume has to be obtained
from a general intermediate size situation. This was shown to be
the case in Ref.~\cite{Cley60}, where expanding
$Z_{S=0}(T,V,\lambda_B)$ of Eq.~(\ref{canon3}) in power series, it
was expressed as $Z_{S=0}=Z_0I_0(x_1)$, where $Z_0$ is the
partition function that includes all particles and resonances with
$S=0$, $I_0(x_1)$ is the modified Bessel function, and
$x_1=2\sqrt{N_{S=1}N_{S=-1}}$. The computed $K^-/K^+$ ratio gave
precisely the same expression as those given here for small
[Eq.(\ref{eq:smallvolume})] and large [Eq.(\ref{eq:largevolume})]
volumes. Therefore, as noted in Ref.~\cite{Cley59}, the
$K^-$/$K^+$ ratio is  independent of the volume and, consequently,
independent on whether it is calculated in the canonical or
grand-canonical schemes.

\subsection{In-medium effects in \mbox{${\boldmath K^-/K^+}$} ratio at finite T}

In this subsection we study how the in-medium modifications of the
properties of the hadrons at finite temperature affect the value of
the $K^-$/$K^+$ ratio, focusing our attention on the properties of the
antikaons in hot and dense matter. For consistency with previous
papers, we prefer to compute the inverse ratio $K^+$/$K^-$. As it was
mentioned before, the number of $K^-$ ($K^+$) has to be balanced by
particles and resonances with $S=+1$($S=-1$) in order to conserve
strangeness exactly. For balancing the number of $K^+$, the main
contribution in the $S=-1$ sector comes from the $\Lambda$ and
$\Sigma$ hyperons and, in a smaller proportion, from the $K^-$ mesons.
In addition, the effect of the $\Sigma^*(1385)$ resonance is also
considered because it is comparable to that of the $K^-$ mesons. On
the other hand, the number of $K^-$ is balanced only by the presence
of $K^+$. Then, we can write the $K^+$/$K^-$ ratio as,

\begin{eqnarray}
\frac{K^+}{K^-} \equiv \frac{N_{K^+}}{N_{K^-}}
=\frac{Z^1_{K^+}(Z^1_{K^-}+Z^1_{\Lambda}+
Z^1_{\Sigma}+Z^1_{\Sigma^*})}{Z^1_{K^-}Z^1_{K^+}}
=1+\frac{Z^1_{\Lambda}+Z^1_{\Sigma}+Z^1_{\Sigma^*}}{Z^1_{K^-}} \ ,
\label{eq:lamb-sig-k}
\end{eqnarray}
where the $Z$'s indicate the different one-particle partition
functions for $K^-$, $K^+$, $\Lambda$, $\Sigma$ and $\Sigma^*$, and
, for baryons, they now contain the corresponding fugacity.
It is clear from Eq.~(\ref{eq:lamb-sig-k}) that the relative abundance
of $\Lambda$, $\Sigma$ and $\Sigma^*$ baryons with respect to that of
$K^-$ mesons determines the value of the ratio.

In order to introduce the in-medium and temperature effects, the
particles involved in the calculation of the ratio are dressed
according to their properties in the hot and dense medium in which
they are embedded. For the $\Lambda$ and $\Sigma$ hyperons, the
partition function
\begin{eqnarray}
Z_{\Lambda,\Sigma}=  g_{\Lambda,\Sigma}  V  \int
\frac{d^3p}{(2\pi)^3} e^{\frac{-E_{\Lambda,\Sigma}+\mu_{N}}{T}},
\label{eq:lam-sig}
\end{eqnarray}
is constructed using a mean-field dispersion relation for the
single-particle energies
\begin{eqnarray}
E_{\Lambda,\Sigma}=\sqrt{m_{\Lambda,\Sigma}^2+
p^2}+U_{\Lambda,\Sigma}(\rho) \ . \label{eq:otros-dressing}
\end{eqnarray}
For $U_{\Lambda}(\rho)$, we take the parameterisation of
Ref.~\cite{Gal97}, $U_{\Lambda}(\rho)=-340\rho+1087.5\rho^2$.  For
$U_{\Sigma}(\rho)$, we take a repulsive potential,
$U_{\Sigma}(\rho)=30 \rho/\rho_0$, extracted from analysis of
$\Sigma$-atoms and $\Sigma$-nucleus scattering
\cite{Mares95,Dabrowski}, where $\rho_0 =0.17$ fm$^{-3}$ is the
saturation density of symmetric nuclear matter.  A repulsive $\Sigma$
potential is compatible with the absence of any bound state or narrow
peaks in the continuum in a recent $\Sigma$-hypernuclear search done
at BNL \cite{Bart}. The $\Sigma^*(1385)$ resonance is described by a
Breit-Wigner shape,
\begin{eqnarray}
Z_{\Sigma*}&=& g_{\Sigma^*} V \int \frac{d^3p}{(2\pi)^3}
\int_{m_{\Sigma^*}-2\Gamma}^{m_{\Sigma^*}+2\Gamma}\ ds \
e^{\frac{-\sqrt{p^2+s}}{T}} \frac{1}{\pi}
\frac{m_{\Sigma^*}\Gamma}{(s-m_{\Sigma^*}^2)^2+m_{\Sigma^*}^2\Gamma^2}
\  e^{\frac{\mu_N}{T}}  \ ,
\label{eq:resonance}
\end{eqnarray}
with $m_{\Sigma^*}=1385 \ {\rm MeV}$ and $\Gamma=37 \ {\rm MeV}$.

In the case of $K^+$ we take
\begin{eqnarray}
Z_{K^+}&=& g_{K^+} V \int \frac{d^3p}{(2\pi)^3} e^{\frac{-E_{K^+}}{T}} \\
E_{K^+}&=&\sqrt{m_{K^+}^2+p^2}+U_{K^+}(\rho) \ ,
\end{eqnarray}
where $U_{K^+}(\rho)= 32 \rho/\rho_0$ is obtained from a $t\rho$
approximation, as discussed in Refs.\cite{weise,Oset-Ramos}.

A particular effort has been invested in studying the antikaon
properties in the medium since the ${\bar K}N$ has a particularly rich
structure due to the presence of the $\Lambda(1405)$ resonance
\cite{Lutz98,Oset00,Schaffner,Laura01}. The antikaon optical potential
in hot and dense nuclear matter has recently been obtained
\cite{Laura02} within the framework of a coupled-channel
self-consistent calculation taking, as bare meson-baryon interaction,
the meson-exchange potential of the J\"ulich group ~\cite{holinde90}.
In order to understand the influence of the in-medium antikaon
properties on the $K^+$/$K^-$ ratio, two different prescriptions for
the single-particle energy of the antikaons have been considered.

First, the so-called on-shell approximation to the antikaon
single-particle energy has been adopted. The antikaon partition
function in this approach reads
\begin{eqnarray}
Z_{K^-}&=& g_{K^-} V \int\frac{d^3p}{(2\pi)^3}
e^{\frac{-E_{K^-}}{T}} \nonumber \\
E_{K^-}&=&\sqrt{m_{K^-}^2 +p^2}+U_{\bar{K}}(T,\rho,E_{K^-},p) \ ,
\label{eq:onshell}
\end{eqnarray}
where $U_{\bar{K}}(T,\rho,E_{K^-},p)$ is the $\bar{K}$ single-particle
potential in the Brueckner-Hartree-Fock approach given by
\begin{equation}
U_{\bar K}(T,\rho,E_{K^-},p)= {\rm Re} \int  \ d^3k \  n(k,T)
\langle \bar K N \mid G_{\bar K N\rightarrow \bar K N}
(\Omega=E_N+E_{\bar{K}},T) \mid \bar K N \rangle \ , \label{eq:self0}
\end{equation}
which is built from a self-consistent effective $\bar{K}N$ interaction
in nuclear symmetric matter, averaging over the occupied nucleonic
states according to the Fermi distribution at a given temperature,
$n(k,T)$.

The second approach incorporates the complete energy-and-momentum
dependent $\bar{K}$ self-energy
\begin{equation}
\Pi_{\bar{K}}(T,\rho,\omega,p)=2 \
\sqrt{p^2+m_{\bar{K}}^2} \
U_{\bar{K}}(T,\rho,\omega,p) \ ,
\label{eq:pik}
\end{equation}
via the corresponding $\bar{K}$ spectral density
\begin{equation}
S_{\bar K}(T,\rho,\omega,p) = - \frac {1}{\pi} {\mathrm Im\,}
D_{\bar K}(T,\rho,\omega,p) \ ,
\label{eq:spec}
\end{equation}
where
\begin{equation}
 D_{\bar{K}}(T,\rho,\omega,p)=
\frac{1}{\omega^2-p^2-m_{\bar{K}}^2-
\Pi_{\bar{K}}(T,\rho,\omega,p)} 
\end{equation}
stands for the $\bar{K}$ propagator. In this case, the $\bar{K}$
partition function reads
\begin{eqnarray}
Z_{K^-}&=& g_{K^-} V \int \frac{d^3p}{(2\pi)^3} \int ds
S_{\bar K}(T,\rho,\omega=\sqrt{s},p) \ e^{\frac{-\sqrt{s}}{T}} \ ,
\label{eq:offshell}
\end{eqnarray}
where $s=\omega^2$.

We note, however, that only the s-wave contribution of the J\"ulich
$\bar{K} N $ interaction has been kept. The reason is that this
potential presents some short-comings in the $L=1$ partial wave, which
manifest themselves especially in the low energy region of the
$\bar{K}$ self-energy.  Specifically, the $\Lambda$ and $\Sigma$ poles
of the ${\bar K}N$ $T$-matrix come out lower about $100 \ {\rm MeV}$
than the physical values and, consequently, the corresponding strength
in the antikaon spectral function due to hyperon-hole excitations
appears at too low energies, a region very important for the
calculation we are conducting here. In addition, the role of the
$\Sigma^*(1385)$ pole, which lies below the $\bar{K}N$ threshold, is
not included in the J\"ulich ${\bar K}N$ interaction. In order to
overcome these problems, we have added to our ${\bar K}$ self-energy
the p-wave contribution as calculated in Ref.~\cite{Oset-Ramos}. In
this model, the p-wave self-energy comes from the coupling of the
$\bar{K}$ meson to hyperon-hole ($YN^{-1}$) excitations, where $Y$
stands for $\Lambda$, $\Sigma$ and $\Sigma^*$. In symmetric nuclear
matter at $T=0$, this self-energy reads
\begin{eqnarray}
\Pi_{\bar{K}}^p(\rho,\omega,\vec{p})& =&\frac{1}{2}
\left(\frac{g_{N \Lambda \bar{K}}}{2 M} \right)^2 \vec{p} \,^2
f_{\Lambda}^2 {\mathcal U}_{\Lambda}(\rho,\omega,\vec{p}) \nonumber \\ &+&
\frac{3}{2} \left( \frac{g_{N \Sigma \bar{K}}}{2 M} \right)^2
\vec{p}\, ^2 f_{\Sigma}^2 {\mathcal U}_{\Sigma}(\rho, \omega,\vec{p})
\nonumber \\ &+& \frac{1}{2} \left( \frac{g_{N \Sigma^*
\bar{K}}}{2 M} \right)^2 \vec{p}\, ^2 f_{\Sigma^*}^2
{\mathcal U}_{\Sigma^*}(\rho,\omega,\vec{p}) \ .
\label{eq:self}
\end{eqnarray}
The quantities $g_{N \Lambda \bar{K}}$, $g_{N \Sigma \bar{K}}$ and
$g_{N \Sigma^* \bar{K}}$ are the $N \Lambda \bar{K}$, $N \Sigma
\bar{K}$ and $N \Sigma^* \bar{K}$ coupling constants, while
$f_{\Lambda}$, $f_\Sigma$, $f_{\Sigma^*}$ are the $\Lambda$, $\Sigma$,
$\Sigma^*$ relativistic recoil vertex corrections and ${\mathcal
  U}_{\Lambda}$, ${\mathcal U}_\Sigma$, ${\mathcal U}_{\Sigma^*}$ the
Lindhard functions at $T=0$. For more details, see
Ref.\cite{Oset-Ramos}. The relevance of this low energy region makes
advisable to extend this p-wave contribution to finite temperature.

The hyperon-hole Lindhard functions at finite temperature are easily
obtained from the $\Delta$-hole one given in Eq.~(A.16) of
Ref.~\cite{Laura02}, by ignoring, due to strangeness conservation, the
crossed-term contribution. The spin-isospin degeneracy factors and
coupling constants need to be accommodated to the notation used in
Eq.~(\ref{eq:self}) and this amounts to replace $\frac{2}{3}
\frac{f^*_{\Delta}}{f_N}$ by $\frac{3}{2}$. Finally, the width
$\Gamma$ in Eq.~(A.16) of Ref.~\cite{Laura02} is taken to zero, not
only for the stable $\Lambda$ and $\Sigma$ hyperons but also, for
simplicity, for the $\Sigma^*$ hyperon, which allows one to obtain the
imaginary part of the Lindhard function analytically.

\section{Results}
\label{sec:results}

In this section we discuss the effects of dressing the ${K^-}$ mesons
in hot and dense nuclear matter on the ${K^-}$/$K^+$ ratio around the
value found in Ni+Ni collisions at an energy of 1.93 $\rm AGeV$. A
preliminary study was already reported in Ref.~\cite{Lauraproceed}.
As previously mentioned, we prefer to discuss the results for the
inverted ratio ${K^+}$/${K^-}$.

\begin{figure}[htb]
  \centerline{ \includegraphics[width=0.6\textwidth,angle=-90]{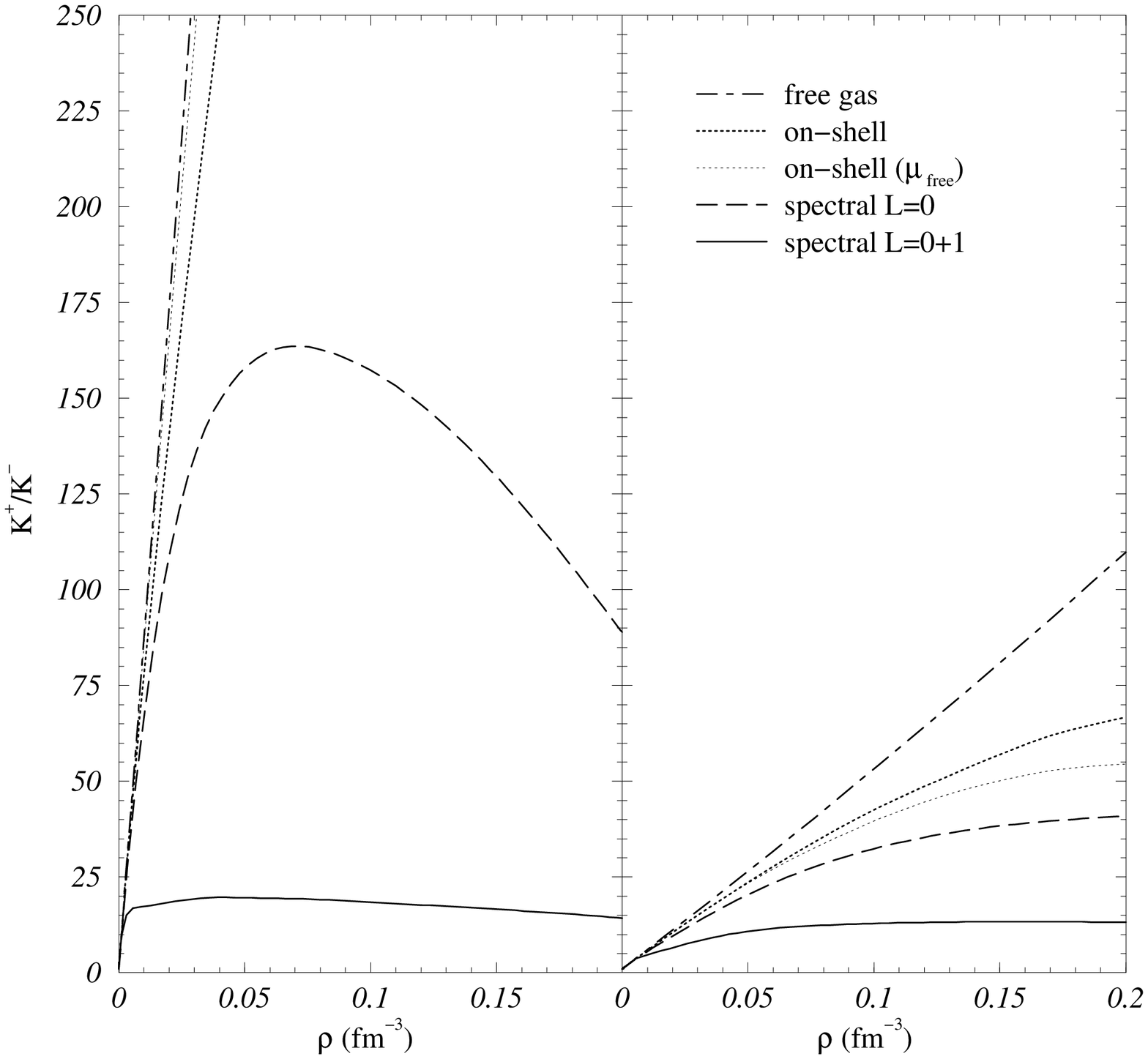} }
      \caption{\small $K^+/K^-$ ratio as a function of the density for $T=50$
        ${\rm MeV}$ (left panel) and $T=80$ ${\rm MeV}$ (right panel)
        calculated in different approaches: the free Fermi gas
        (dot-dashed line), on-shell self-energies (dotted line),
        on-shell self-energies with $\mu$ from a free
        (non-interacting) Fermi gas (thin dotted line), dressing the
        $K^-$ with its single particle spectral function, with the
        $L=0$ contribution (long-dashed line) and taking into account the
        additional $L=1$ partial wave (solid line).}
 
        \label{fig:ratio1}
\end{figure}

The ${K^+}$/$K^-$ ratio is shown in Fig.~\ref{fig:ratio1} as a
function of density at two given temperatures, T= 50 and 80 MeV,
calculated for the three ways of dressing of the $K^-$: free
(dot-dashed lines), the on-shell or mean-field approximation of
Eq.~(\ref{eq:onshell}) (dotted lines), and using the $\bar K$ spectral
density including s-wave (long-dashed lines) or both s-wave and p-wave
contributions (solid lines). The two chosen temperatures roughly
delimit the range of temperatures which have been claimed to
reproduce, in the framework of the thermal model, not only the
${K^+}$/$K^-$ ratio but also all the other particle ratios involved in
the Ni+Ni collisions at SIS energies\cite{Cley57,Cley59}.

Since the baryonic chemical potential $\mu_B$ grows with density, the
factor $e^{\mu_B/T}$ in the partition functions of
Eqs.~(\ref{eq:lam-sig}), (\ref{eq:resonance}) allows one to understand
why the ratio increases so strongly with density in the free gas
approximation (dot-dashed lines).  The same is true when the particles
are dressed. In this case, however, the $K^-$ feels an increasing
attraction with density which tends to compensate the variation of
$\mu_B$ and the curves bend down after the initial increase.  This
effect is particularly notorious when the full $\bar K$ spectral
density is used.  The results are in qualitative agreement with the
``broad-band equilibration" notion introduced by Brown et al.
\cite{brown01}.  However, in this present model, the gain in binding
energy in the on-shell approximation for $K^-$ (thick dotted line in
Fig.~\ref{fig:ratio1}) when the density grows does not completely
compensate the increase of $\mu_B$, as was the case in
Ref.~\cite{brown01}. To illustrate this fact we note that the
variation of the $K^-$ single particle energy at zero momentum at
$T=70$ MeV changes in our model \cite{Laura02} from 434 MeV to 375 MeV
when the density grows from 1.2$\rho_0$ to 2.1$\rho_0$, while $\mu_B$
changes from 873 MeV to 962 MeV.  Therefore the relevant quantity to
understand the behaviour of the $K^+$/$K^-$ ratio with density in the
on-shell approximation, i.e. the sum of $\mu_B$ and $E_{K^-}$ [see
Eq.~(6) of Ref.~\cite{brown01}], suffers in our model a variation of
about 30 MeV.  On the other hand, the model of Ref.\cite{brown01}
assumed a slower variation of $\mu_B$, from 860 MeV to 905 MeV, which
was almost cancelled by the change of $E_{K^-}$ from 380 MeV to 332
MeV, giving therefore a practically density independent ratio. We note
that our chemical potential is derived from a nucleonic energy
spectrum obtained in a Walecka $\sigma-\omega$ model, using density
dependent scalar and vector coupling constants fitted to reproduce
Dirac-Brueckner-Hartree-Fock calculation [see Table 10.9 in
Ref.\cite{Mach89}].

The results obtained in the present microscopic calculation show that
the ``broad-band equilibration" only shows up clearly when the full
spectral function is used (solid line in Fig.~\ref{fig:ratio1}). After
an increase at low densities, the $K^+/K^-$ ratio remains constant at
intermediate and high densities. The use of the spectral density
implicitly amounts to an additional gain in binding energy for the
antikaons and, as density increases, it compensates rather well the
variation of $\mu_B$.

\begin{figure}[htb]
  \centerline{
    \includegraphics[width=0.6\textwidth,angle=-90]{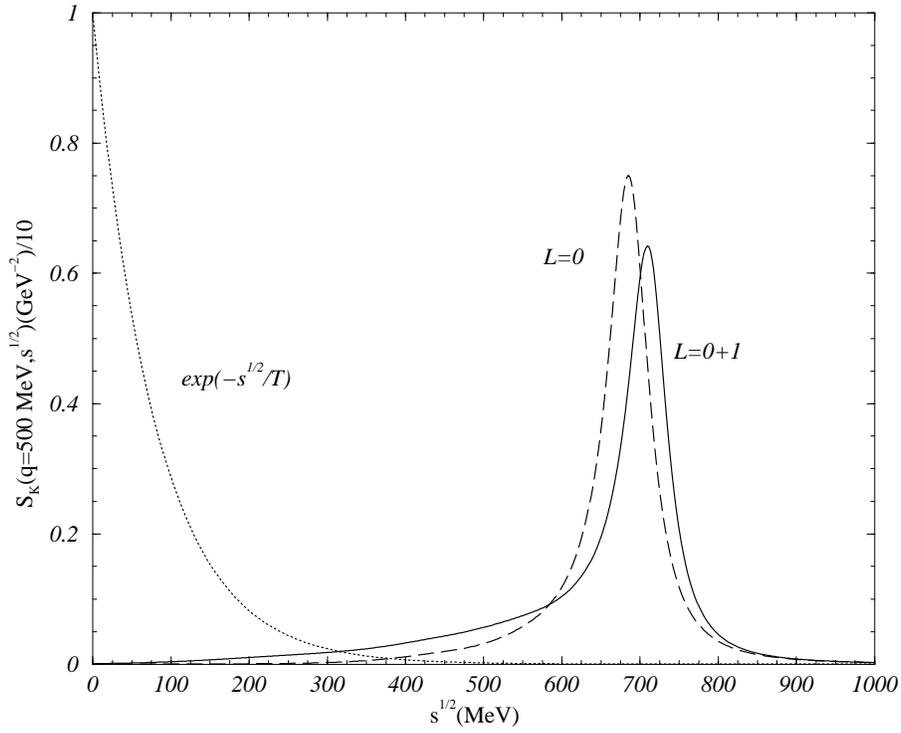} }
      \caption{\small  The Boltzmann factor (dotted line) and the $K^-$ spectral
        function, including s-wave (dashed line) or s- and p-wave
        (solid line) components of the ${\bar K}N$ interaction, as
        functions of the energy, for a momentum $q=500$ MeV at
        saturation density and temperature $T= 80$ MeV. }
        \label{fig:ratio2}
\end{figure}

To understand the origin of this additional effective attraction when
the full spectral density is used, we show in Fig.~\ref{fig:ratio2}
the two functions that contribute to the integral over the energy in
the definition of the $K^-$ partition function
[Eq.(\ref{eq:offshell})], namely the Boltzmann factor
$e^{-\sqrt{s}/T}$ and the $K^-$ spectral functions including L=0
(long-dashed line) and L=0+1 (solid line) components of the ${\bar
  K}N$ interaction, for a momentum $q=500$ MeV at $\rho=0.17$
fm$^{-3}$ and T$=80$ MeV.  As it is clearly seen in the figure, the
overlap of the Boltzmann factor with the quasi-particle peak of the
$K^-$ spectral function is small for this momentum.  It is precisely
the overlap with the strength in the low energy region that acts as a
source of attraction in the contribution to the $K^-$ partition
function. This effect is particularly pronounced when the p-waves are
included, due to the additional low-energy components in the spectral
function coming from the coupling of the $K^-$ meson to hyperon-hole
($YN^{-1}$) excitations, where $Y$ stands for $\Lambda$, $\Sigma$ and
$\Sigma^*$. Assigning these low energy components to real antikaons in
the medium is not clear, since one should interpret them as
representing the production of hyperons through $\bar{K}N \rightarrow
Y$ conversion. While this is certainly true, it may also happen that,
once these additional hyperons are present in the system, they can
subsequently interact with fast non-strange particles (pions,
nucleons) to create new antikaons. A clear interpretation on what
fraction of the low energy strength will emerge as antikaons at
freeze-out is certainly an interesting question and its investigation
will be left here for forthcoming work.

\begin{figure}[htb]
  \centerline{
    \includegraphics[width=0.6\textwidth,angle=-90]{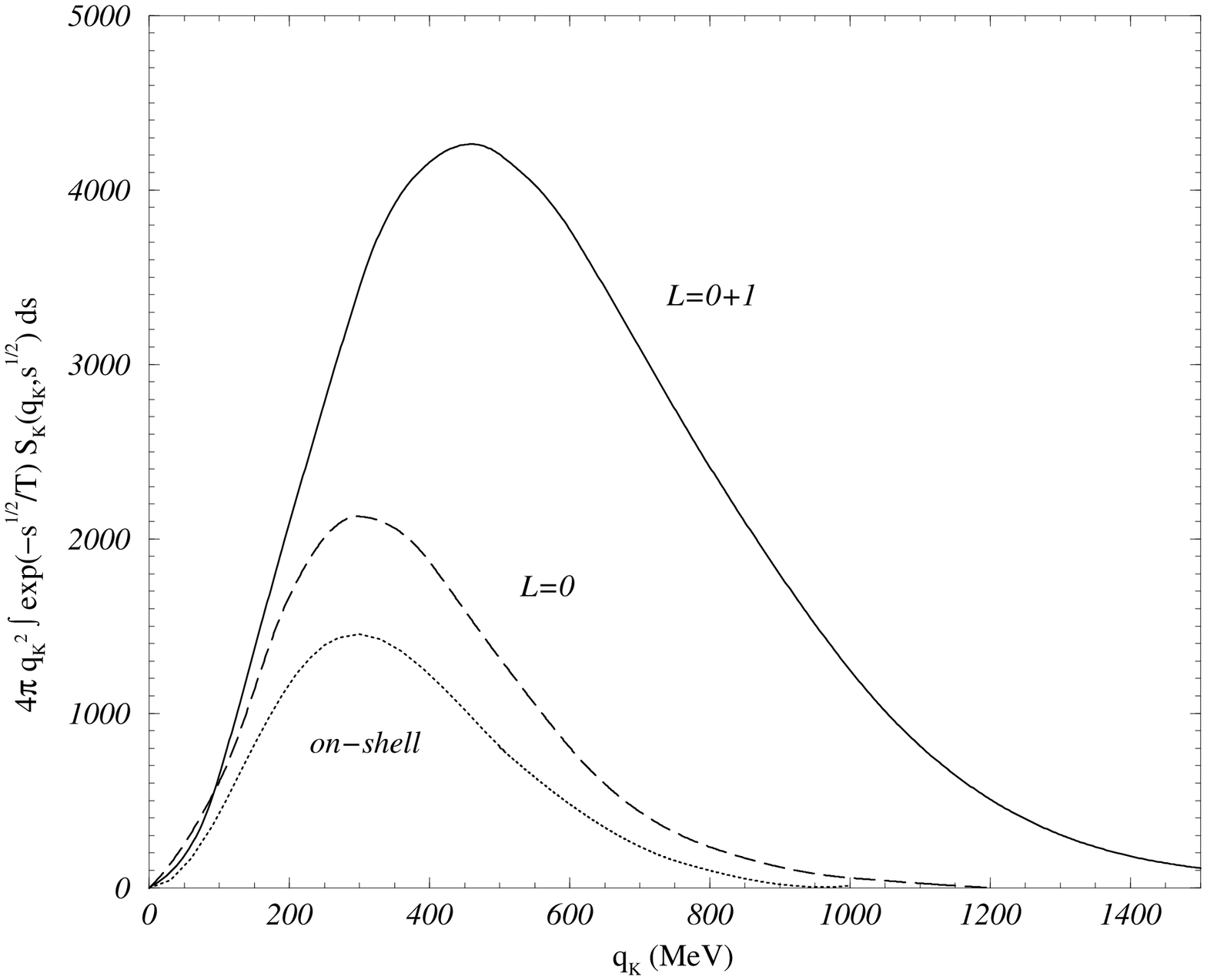} }
      \caption{\small The integrand which defines the $K^-$ partition function
        [Eq.~(\ref{eq:offshell})] as a function of momentum, at
        saturation density and $T=80$ MeV, for different approaches:
        on-shell prescription (dotted line), using the $K^-$ spectral
        function with the L=0 components of the ${\bar K}N$
        interaction (long-dashed line) and including also the L=1 partial
        waves (solid line). }
        \label{fig:ratio3}
\end{figure}

Once the integral over the energy is performed, the determination of
the $K^-$ partition function still requires an integral over the
momentum. The integrand as a function of momentum is plotted in
Fig.~\ref{fig:ratio3} for the same density and temperature than in the
previous figure.  As expected, the integrand is larger when the full
spectral density is considered. In this case, the $K^-$ partition
function is enhanced and therefore the ${K^+}$/$K^-$ ratio is smaller
than in the on-shell approximation and also in the case where only the
L=0 contributions to the spectral density are used. Notice the
behaviour at large $q$, which decays very quickly in the on-shell
approximation but has a long tail for the $L=0+1$ spectral density
originated from the coupling of the $K^-$ meson to $Y N^{-1}$
configurations.

\begin{figure}[htb]
  \centerline{ \includegraphics[width=0.6\textwidth,angle=-90]{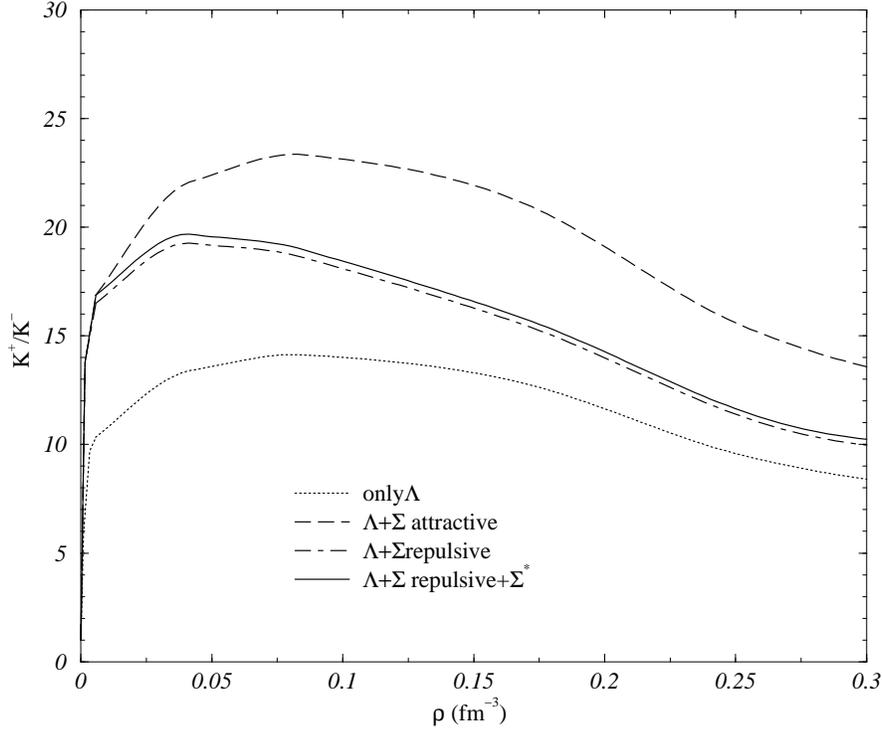} }
      \caption{\small The $K^+$/$K^-$ ratio as a function of density at
        T=50 MeV. The dotted line shows the results when only the
        $\Lambda$ hyperons are considered in the determination of the
        ratio. The dashed (dot-dashed) line includes also the
        contribution of the $\Sigma$ hyperon dressed with an
        attractive (repulsive) mean-field potential. The solid line
        includes the effect of the $\Sigma^*$ resonance.}
 
\label{fig:ratio4}
\end{figure}

Another aspect that we want to consider is how the dressing of the
$\Sigma$ hyperon affects the value of the ratio.  In
Fig.~\ref{fig:ratio4} the value of the ${K^+}$/$K^-$ ratio at $T=50$
MeV is shown as a function of density for different situations. In all
calculations displayed in the figure, the partition function
associated to $K^-$ has been obtained using the full $K^-$ spectral
density. The dotted line corresponds to the case where only the
$\Lambda$ hyperons, dressed with the attractive mean-field potential
given in the previous section, are included to balance strangeness.
When the $\Sigma$ hyperon is incorporated with a moderately attractive
potential of the type $U_{\Sigma}=-30\rho/\rho_0$ MeV, the $K^+$/$K^-$
ratio is enhanced substantially (dashed line). This enhancement is
more moderate when one uses the repulsive potential
$U_{\Sigma}=+30\rho/\rho_0$ instead (dot-dashed line).  Finally, the
additional contribution of the $\Sigma^*$ resonance produces only a
small increase of the ratio (solid line) due to its higher mass. We
have checked that heavier strange baryonic or mesonic resonances do
not produce visible changes in our results.  Notice also that,
although the ratios obtained with both prescriptions for the
mean-field potential of the $\Sigma$ meson differ appreciably, the
present uncertainties in the ratio would not permit to discriminate
between them.
 
\begin{figure}
  \centerline{ \includegraphics[width=0.6\textwidth]{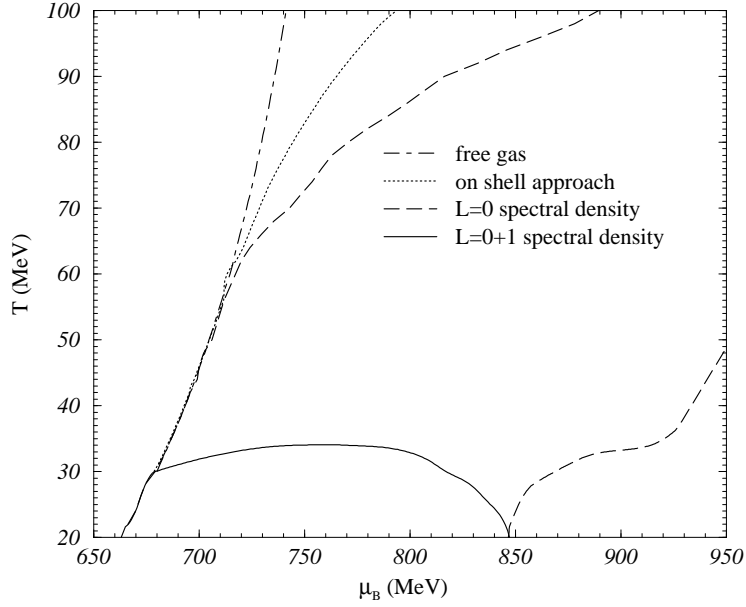} }
  \caption{\small Relation between the temperature and the
    baryochemical potential of hadronic matter produced in heavy-ion
    collisions for fixed $K^+/K^-$ ratio of 30, calculated within
    different approaches as discussed in the text.}
\label{fig:ratio6}
\end{figure}

In the framework of the statistical model, one obtains a relation
between the temperature and the chemical potential of the hadronic
matter produced in the heavy-ion collisions by fixing the value of the
${K^-}$/$K^+$ ratio which was measured for Ni+Ni collisions at GSI to
be on the average ${K^-}$/$K^+ = 0.031\pm 0.005$ \cite{menzel}.  We
compare our results with a corresponding inverse ratio of
${K^+}$/$K^-=30$ in the following.  The temperatures and chemical
potentials compatible with that ratio are shown in
Fig.~\ref{fig:ratio6} for different approaches.  The dot-dashed line
stands for a free gas of hadrons, similar to the calculations reported
in Refs.~\cite{Cley57,Cley59}. The dotted line shows the $T(\mu)$
curve obtained with the on-shell or mean-field approximation [see
Eqs.(\ref{eq:onshell}) and (\ref{eq:self0})], while the dashed and
solid lines correspond to the inclusion of the off-shell properties of
the ${K^-}$ self-energy by using its spectral density [Eqs.
(\ref{eq:pik}),(\ref{eq:spec}) and (\ref{eq:offshell})], including L=0
or L=0+1 components, respectively.

In the free gas limit, the temperatures compatible with a ratio
$K^+$/${K^-}=30$ imply a narrow range of values for the baryonic
chemical potentials, namely $\mu_B \in [665,740]$ $\rm MeV$ for
temperatures in the range of 20 to 100 MeV. These values translate
into density ranges of $\rho \in [6 \times 10^{-7}\rho_0, 0.9\rho_0
]$.

As it can be seen from the dotted line in Fig.~\ref{fig:ratio6}, the
attractive mean-field potential of the antikaons compensates the
effect of increasing baryochemical potential $\mu_B$. As a
consequence, the density at which the freeze-out temperature
compatible with the measured ratio takes place also grows.  But this
attraction is not enough to get the same $K^+$/$K^-$ ratio for a
substantially broader range of density compared to the free case. So
we do not see a clear indication of ``broad-band equilibration'' in
our self-consistent mean-field calculation in contrast to the results
of Brown, Rho, and Song \cite{brown01}.

The influence of the antikaon dressing on the ratio is much more
evident when the spectral density is employed (long-dashed and solid
lines).  From the preceding discussions, it is easy to understand that
the low energy behaviour of the spectral density enhances the $K^-$
contribution to the ratio, having a similar role as an attractive
potential and, hence, the value of $\mu_B$ compatible with a ratio at
a given temperature increases.  Moreover, due to the bending of the
$K^+$/$K^-$ ratio with density and its evolution with temperature
observed in Fig.~\ref{fig:ratio1}, it is clear that there will be a
maximum value of T compatible with a given value of the ratio. Below
this maximum temperature, there will be two densities or,
equivalently, two chemical potentials compatible with the ratio. For
example, the ratio $K^+$/$K^-$=30 will in fact not be realized with
the temperatures of $T=50$ MeV and $T=80$ MeV displayed in
Fig.~\ref{fig:ratio1}, if the antikaon is dressed with the spectral
density containing $L=0$ and $L=1$ components.  As shown in
Fig.~\ref{fig:ratio6}, only temperatures lower than $34$ MeV are
compatible with ratio values of 30!

We note that the flat regions depicted by the solid lines in
Fig.~\ref{fig:ratio6} could be considered to be in correspondence with
the notion of ``broad-band equilibration" of Brown et al.
\cite{brown01}, in the sense that a narrow range of temperatures and a
wide range of densities are compatible with a particular value of the
$K^+$/$K^-$ ratio.  Nevertheless, the temperature range is too low to
be compatible with the measured slope parameter of the pion spectra.
Explicitly, for $K^+$/$K^-$=30, we observe a nearly constant ratio
observed in the range of $30-34$ MeV covering a range of chemical
potentials in between $680-815$ MeV which translates into a density
range $\rho \in [1.5 \times 10^{-4}\rho_0, 0.02\rho_0]$. Note that in
this case, we can hardly speak of a broad-band equilibration in the
sense of that introduced by Brown, Rho and Song in
Ref.~\cite{brown01}, where a ratio $K^+$/$K^-$=30 holds over a large
range of densities in between $\frac{1}{4}\rho_0$ and $2\rho_0$ for
$T=70$ MeV.  However, as we indicated at the beginning of this
section, this result was obtained in the framework of a mean-field
model and our equivalent on-shell results (dashed lines in
Figs.~\ref{fig:ratio1}, \ref{fig:ratio6}), based on a stronger
variation of the $\mu_B$ with density and on a less attractive
$U_{\bar K}$, seem to be very far from producing the broad-band
equilibration behaviour.

As pointed out before, our nucleon chemical potential is obtained in
the framework of a relativistic model and varies with density more
strongly than that used in Ref.~\cite{brown01}, which shows values
close to those for a free Fermi gas. If we now calculate the
$K^+$/$K^-$ ratio using a $\mu_B(\rho,T)$ for a free (non-interacting)
system in the on-shell approximation, we obtain the thin dotted line
in Fig.~\ref{fig:ratio1}. At $T=80$ MeV, we now observe a tendency of a
broad-band equilibration but for ratios higher than $30$, of around
$50$. This is connected to the particular on-shell potential of the
antikaon which, in our self-consistent procedure, turns to be
moderately attractive. Only if the attraction was larger would the
broad band be realized in this on-shell picture for smaller values of
the ratio as found by Brown et al. \cite{brown01}.

\begin{figure}
  \centerline{ \includegraphics[width=0.6\textwidth]{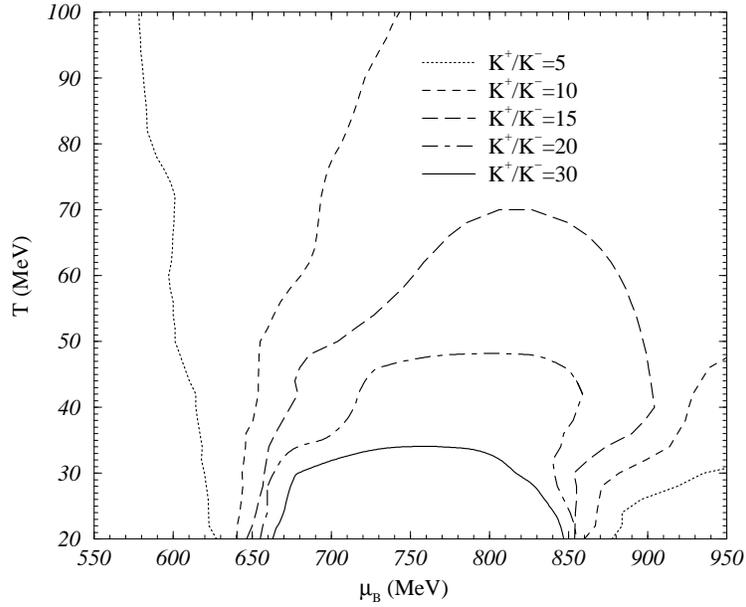} }
  \caption{\small The $K^+/K^-$ ratio plotted for the full spectral
    density of the $K^-$ as a contour plot for different temperatures
    and baryochemical potentials.}
\label{fig:ratio7}
\end{figure}

Fig.~\ref{fig:ratio7} shows the $K^+$/$K^-$ ratio for the full model
calculation as a contour plot for different temperatures and
baryochemical potentials. We note, that the ratio is substantially
lower at the temperature and density range of interest, being more
likely around 15 or so for a moderately large region of baryochemical
potential. Note that this reduced ratio translates into an overall
enhanced production of $K^-$ by a factor two compared to the
experimentally measured value. At pion freeze-out, the medium can hold
twice as many $K^-$ as needed to explain the measured enhanced
production of $K^-$ if one considers the full spectral features of the
$K^-$ in the medium. We stress again that this enhancement is not due
to an increased attraction in the sense of a mean-field calculation.
It is a consequence of the additional strength of the spectral
function at low energies which emerges when taking into account p-wave
hyperon-hole excitations. The Boltzmann-factor amplifies the
contribution of the low-energy region of the spectral function so that
these excitations are becoming the main reason for the overall
enhanced production of the $K^-$ in the medium.

\section{Conclusions}
\label{sec:conclusions}

We have studied, within the framework of a statistical model, the
influence of considering the modified properties in hot and dense
matter of the hadrons involved in the determination of the $K^-/K^+$
ratio produced in heavy-ion collisions at SIS/GSI.  We have focused
our attention on incorporating the effects of the antikaon
self-energy, which was derived within the framework of a
self-consistent coupled-channel calculation taking, as bare
interaction, the meson-exchange potential of the J\"ulich group for
the s-wave and adding the p-wave components of the $Yh$ excitations,
with $Y=\Lambda,\Sigma,\Sigma^*$.

It is found that the determination of the temperature and chemical
potential at freeze-out conditions compatible with the ratio $K^-/K^+$
is very delicate and depends very strongly on the approximation
adopted for the antikaon self-energy. The effect of dressing the $K^-$
with an spectral function including both s- and p-wave components of
the ${\bar K}N$ interaction lowers considerably the effective
temperature while increasing the chemical potential $\mu_B$ up to
$850$ MeV, compared to calculations for a noninteracting hadron gas.

When the free or on-shell properties of the antikaon are considered,
the ratio at a given temperature shows a strong dependence with the
density. This is in contrast with the ``broad-band equilibration''
advocated by Brown et al. \cite{brown01}, which was established, in
the context of a mean-field picture, through a compensation between
the increased attraction of the mean-field $\bar{K}$ potential as
density grows with the increase in the baryon chemical potential.
Our mean-field properties do not achieve such a compensation due to a
stronger increase of the nucleon chemical potential $\mu_B$ with
density. 

When taking into account the full features of the spectral function of
the $K^-$ via hyperon-hole excitations, we find that the $K^-/K^+$
ratio exhibits broad-band equilibration. Nevertheless, the ratio is
even in excess of the measured ratio at temperatures which are deduced
from the slope parameter for pions. One can argue in principle, that
dynamical non-equilibrium effects can reduce the number of $K^-$ by
virtue of annihilation with nucleons to hyperons and pions at
freeze-out. The number of the latter produced hadrons will not be
altered substantially by this annihilation process as the $K^-$ is much
less abundant than hyperons and especially pions. In our approach, we
calculate the number density or particle ratios which are in the
medium. What needs to be clarified is how the particles get on-shell
at freeze-out, a question previously posed e.g.\ for antiproton and
antihyperon production at the GSI. Clearly, this has to be addressed
in dynamical models and we leave that investigation for forthcoming
work.

\section*{Acknowledgements}
We are very grateful to Dr.~Volker Koch and Dr.~Luis Alvarez-Ruso for
useful discussions.  This work is partially supported by DGICYT
project BFM2001-01868 and by the Generalitat de Catalunya project
2001SGR00064. L.T. also wishes to acknowledge support from the
Ministerio de Educaci\'on y Cultura (Spain).

\newpage \renewcommand{\theequation}{\Alph{section}.\arabic{equation}}




\end{document}